# A New Trusted and Secured E-commerce Architeture for Cloud Computing

Kawser Wazed Nafi, Tonny Shekha Kar, Md. Amjad Hossain, M. M. A. Hashem
Computer Science and Engineering Department,
Khulna University of Engineering and Technology
Khulna, Bangladesh
kwnafi@yahoo.com, tulip0707051@gmail.com, amjad_kuet@yahoo.com, mma.hashem@outlook.com

*Abstract*—Cloud computing platform gives people the opportunity for sharing resources, services and information among the people of the whole world. In private cloud system, information is shared among the persons who are in that cloud. Presently, different types of internet based systems are running in Cloud Computing environment. E-commerce is one of them. Present models are not secured enough for executing e-transactions easily, especially in cloud platform. Again, most of the time, clients fail to distinguish between the good online business companies and the bad one, which discourages clients and companies to migrate in cloud. In this paper, we have proposed a newer e-commerce architecture depends on encryption based secured and fuzzy logic based certain trust model which will be helpful to solve present e-commerce problems. We had discussed about the whole working procedure of the model in this paper. Finally, at the end of this paper, we have discussed some experimental results about our proposed model which will help to show the validity of our model.

## I. INTRODUCTION

At the present world of networking system, Cloud computing is one the most important and developing concept for both the developers and the users. Persons who are interrelated with the networking environment, cloud computing is a preferable platform for them. Therefore in recent days providing security has become a major challenging issue in cloud computing. Presently, online businesses are very popular to all the people of the world. People want to buy all the things easily and hazard free by sitting in room or any other places and continue business easily. But they face different types of problems at the time of doing e-commerce transactions related to trust and security, like account hacking, miss transaction, intruders' problems, facing problem in finding out right and best business companies, etc.

In the cloud environment, resources are shared among all of the servers, users and individuals. As a result files or data stored in the cloud become open to all. Therefore, data or files of an individual can be handled by all other users of the cloud. [1] Thus the data or files become more vulnerable to attack. As a result it is very easy for an intruder to access, misuse and destroy the original form of data. An intruder can also interrupt the communication. Besides, cloud service providers provide different types of applications which are of very critical nature. Hence, it is extremely essential for the cloud to be secure. Another problem with the cloud system is that an individual may not have control over the place where the data needed to be stored. A cloud user has to use the resource allocation and scheduling, provided by the cloud service provider. Thus, it is also necessary to protect the data or files in the midst of unsecured processing. In order to solve this problem we need to apply security in cloud computing platforms. In our proposed security model we have tried to take into account the various security breaches as much as possible. At present, in the area of cloud computing different security models and algorithms are applied. But, these models have failed to solve all most all the security threats. [2] Moreover for E-commerce [3] and different types of online business, we need to imply high capacity security models in cloud computing fields.

Trust is a well-known concept in everyday life and often serves as a basis for making decisions in complex situations. There are numerous approaches for modeling trust concept in different research fields of computer science, e.g., virtual organizations, mobile and P2P networks, and E-Commerce. The sociologist Diego Gambetta has provided a definition, which is currently shared or at least adopted by many researchers. According to him "Trust is a particular level of the subjective probability with which an agent assesses that another agent or group of agents will perform a particular action, both before he can monitor such action and in a context in which it affects its own action" [4].

The trustworthiness of the overall system depends on the following things:

    a. The trustworthiness of the subsystems and atomic component independently from how these trust values are assessed.
    b. Information on how the system combines its subsystems and components.
    c. The knowledge about which subsystems and components are redundant.

Therefore, a major challenge of serving trust for the overall system is needed to consider that in real world applications the information about the trustworthiness of the subsystems and components itself is subject to uncertainty. Besides, evaluating the models for the trustworthiness of complex systems are needed to be capable of modeling the uncertainty and also calculate and express the degree of uncertainty associated to the derived trustworthiness of the overall system. Trust is interrelated with people's everyday life. When people want to do some new things like selling or buying things from one to other, finding new materials from different source or some other things, concept of trust then arise. In the field of computer science and virtual world of modern technologies like virtual organizations, mobile or P2P networks and E-commerce [5], trust is very important.

In this paper, we have proposed a newer secured and trustworthy e-commerce cloud computing based architecture. We combine fuzzy logic based certain trust model for ensuring trust and encryption algorithm based security model for developing this architecture. Here, two parameters are helpful for calculating trust for a company and then these will help to distinguish between different online business companies of the world. For, secured e-transaction, encryption based security model will be helpful to solve security related problems of present e-commerce system.

This paper is formatted in the following way: - section II describes the related work of trust and security model, section III discusses about the background algorithms and systems of proposed E-commerce architecture, section IV describes the proposed e-commerce architecture, section V discusses about lab's experimental result and section VI describes our future plan about proposed model.

## II. RELATED WORKS

Different ways are used for modeling (un-)certainty of trust values in the field of trust modeling in Cloud computing and internet based marketing sys-tem [6]. But, these models have less capability to derive trustworthiness of a system which are based on knowledge about its components and subsystems. These models help to save from robustness to attacks, e.g. misleading recommendations, Sybil attacks, etc.

Fuzzy logic was used to provide trust in Cloud computing. Different types of attacks and trust models in service oriented systems, distributed system and so on are designed based on fuzzy logic system [7]. But it models different type of uncertainty known as linguistical uncertainty or fuzziness. In paper [8], a very good model for E-commerce, which is based on fuzzy logic, is presented. But, this model also works with uncertain behavior. Belief theory such as Dempster-Shafer theory was used to provide trust in Cloud computing [9]. But the main drawback of this model is that the parameters for belief, disbelief and uncertainty are dependent on each other. It is possible to model uncertainty using Bayesian probabilities [10] which lead to probability density functions e.g., Beta probability density function. It is also possible to apply the propositional standard operators to probability density functions. But this leads to complex mathematical operations and multi-dimensional distributions which are also hard to interpret and to visualize. An enhanced model recently being developed for using in Cloud computing is known as Certain Trust. This model evaluates propositional logic terms that are subject to uncertainty using the propositional logic operators AND, OR and NOT[4].

Numerous researches on security in cloud computing have already been proposed and done in recent time. Identification based cloud computing security model have been worked out by different researchers [11]. But only identifying the actual user does not all the time prevent data hacking or data intruding in the database of cloud environment. Yao's Garbled Circuit is used for secure data saving in cloud servers [12]. It is also an identification based work. The flaw in this system is that it does not ensure security in whole cloud computing platform. Research related to ensuring security in whole cloud computing environments was already worked out in different structures and shaped. AES based file encryption system is used in some of these models [13]. But these models keep both the encryption key and encrypted file in one database server. Only one successful malicious attack in the server may open the whole information files to the hacker, which is not desirable. Some other models and secured architectures are proposed for ensuring security in cloud computing environment [14]. Although these models ensures secured communication between users and servers, but they do not encrypt the loaded information. For best security ensuring process, the uploaded information needs to be encrypted so that none can know about the information and its location. Recently some other secured models for cloud computing environment are also being researched [15]. But, these models also fail to ensure all criteria of cloud computing security issues.

## III. BACKGROUND

Proposed E-commerce architecture combines following two models for ensuring trust and security:-

### A. Certain Trust Model:

Certain Trust Model was constructed for modeling those probabilities, which are subject to uncertainty. This model was designed with a goal of evidence based trust model. Moreover, it has a graphical, intuitively interpretable interface [16] which helps the users to understand the model.

At first a relationship between trust and evidence is needed. For this, they had chosen a Bayesian approach. It is because it provides means for deriving a subjective probability from collected evidence and prior information. At developing a representation of trust, it is necessary to consider to whom trust is represented. It is easy for a software component or a software agent to handle mathematical representations of trust. For it, Bayesian representation of trust is appropriate. The computational model of Certain Trust proposes a new approach for aggregating direct evidence and recommendations. In general, recommendations are collected to increase the amount of information available about the candidates in order to improve the estimate of their trustworthiness. This recommendation system needs to be integrated carefully for the candidates and for the users and owner of cloud servers. This is called robust integration of recommendations. In order to improve the estimate of the trustworthiness of the candidates, it is needed to develop recommendation system carefully. This is called robust integration of recommendations.

Three parameters used in certain logic: average rating t, certainty c, initial expectation f. The average rating t indicates the degree to which past observations support the truth of the proposition. The certainty c indicates the degree to which the average rating is assumed to be representative for the future. The initial expectation f expresses the assumption about the truth of a proposition in absence of evidence.

The equations for these parameters are given below:-

Equation for average rating, $t = \begin{cases} 0.5 & if\ r+s = 0 \\ r/(r+s) & else \end{cases}$ (1)

Here, r represents number of positive evidence and s represents number of negative evidence defined by the users or third person review system.

Equation for certainty, $c = \frac{N(r+s)}{2.w.(N-(r+s))+N.(r+s)}$ (2)

Here, w represents dispositional trust which influences how quickly the final trust value of an entity shifts from base trust value to the relative frequency of positive outcomes and N represents the maximum number of evidence for modeling trust. Using these parameters the expectation value of an opinion E(t,c,f) can be defined as follows:

$$E(t,c,f) = t*c + (1-c)*f \quad (3)$$

The parameters for an opinion o = (t, c, f) can be assessed in the following two ways: direct access and Indirect access. Certain Trust evaluates the logical operators of propositional logic that is AND, OR and NOT. In this model these operators are defined in a way that they are compliant with the evaluation of propositional logic terms in the standard probabilistic approach. However, when combining opinions, those operators will especially take care of the (un)certainty that is assigned to its input parameters and reflect this (un)certainty in the result. The definitions of the operators as defined in the CTM are given in the table 1.

TABLE I. DEFINITION OF OPERATORS

| | |
|---|---|
| OR | $c_{A\vee B} = c_A + c_B - c_A c_B - \frac{cA(1-cB)fB(1-tA)+(1-cA)cBfA(1-tB)}{fA+fB-fAfB}$ <br> $t_{A\vee B} = \begin{cases} \frac{1}{c_{A\vee B}}(c_A t_A + c_B t_B - c_A c_B t_A t_B) & if\ c_{A\vee B} \neq 0 \\ 0.5 & else \end{cases}$ <br> $f_{A\vee B} = f_A + f_B - f_A f_B$ |
| AND | $c_{A\wedge B} = c_A + c_B - c_A c_B - \frac{(1-cA)cB(1-fA)tB+cA(1-cB)(1-fB)tA}{1-fAfB}$ <br> $t_{A\wedge B} = \begin{cases} \frac{1}{c_{A\wedge B}}(c_A c_B t_A t_B + \frac{cA(1-cB)(1-fA)fBtA+(1-cA)cBfA(1-fB)tB}{1-fAfB}) & if\ c_{A\wedge B} \neq 0 \\ 0.5 & else \end{cases}$ <br> $f_{A\wedge B} = f_A f_B$ |
| NOT | $t_{\neg A} = 1 - t_A$, $c_{\neg A} = 1 - c_A$ and $f_{\neg A} = 1 - f_A$ |

For make this model more powerful and helpful for both human and machines, With the help of these parameters and operators derived from certain trust, we have defined two new parameters, Trust T and behavioral probability, P [16]. Trust T is calculated from certainty c and average rating t. the equation is:

Trust, $T = \frac{c*t}{High scaling value of rating} * 100\%$ (4)

Here, High scaling value of rating means the upper value of the range of rating.

Calculating T, we have applied FAM rule of fuzzy logic for creating a relation between certainty c and average rating t. Trust T represents this relation in percentage such a way that the quality of the product can easily be understood. Another parameter, behavioral probability, P, represents how the present behavior of the system varies from its initially expected value and it is proposed with the help of probabilistic logic. It may be less, equal or higher than the initial expectation given by the system developer or the manager of the office. The equation for P is:

Behavioral probability, $P = \frac{(T)-f}{f} * 100\%$ (5)

If T<f, lower probability to show expected behavior
If T>f, higher probability to show expected behavior
If T=f, balanced with the expected behavior

From the equation of P, values with two type magnitude have been found. If it is negative, then it is assumed that it will behave lower than the expected. If it is positive, then higher behavior will be shown by the system than the expected behavior of this system, which is defined by the developer or someone related to the system.

*B. Secured Data Storage and Transferring System:*

Cloud Security model, which is proposed for the present E-commerce architecture, are based on the following security algorithms: RSA, MD5, AES (128 bit key) and OTP(One Time Password) [17]. Figure shows the basic security structure for cloud computing. RSA encryption algorithm is used for making the communication safe. Usually the users' requests are encrypted while sending to the cloud service provider system. System's public key is used for the encryption. Whenever the user requests for a file, the system sends the file by encrypting it via RSA encryption algorithm using the user's public key. Same process is also applied about the user password requests, while logging in the system later. After receiving an encrypted file from the system the user's browser will decrypt it with RSA algorithm using the user's private key. Similarly when the system receives an encrypted file from the user it will immediately decrypt it using its private key. As a result the communication becomes secured between the user and the system.

In the proposed security model one time password is used for authenticating the user. The password is used to keep the user account secure and secret from the unauthorized user. But the user defined password can be compromised. To overcome this difficulty one time password is used in the proposed security model. Thus whenever a user login in the system, he/she will be provided with a new password for using it in the next login. This is usually provided by the system itself. This password will be generated randomly. Each time a new password is created for a user, the previous password for that user will be erased from the system. New password will be updated for that particular user. A single password will be used for login only once. The password will be sent to the users authorized mail account. Therefore at a same time a check to determine the validity of the user is also performed. As a result only authorized user with a valid mail account will be able to connect to the cloud system. By this system, existence of unauthorized user or a user with an invalid mail account will be pointed out. The newly generated password is restored in the system after md5 hashing. The main purpose of MD5 hashing is that this method is a one way system and unbreakable. Therefore it will be difficult for an unauthorized or unknown party for retrieving the password for a selected user even if gained access to the system database.

## IV. PROPOSED E-COMMERCE ARCHITECHTURE

Figure 2 describes the full proposed E-commerce architecture which ensures secured and trustworthy transactions between different elements of E-Commerce architecture. Here, a broker system is introduced between different companies (Like Amazon, E bay or any other e-business companies) which will help to show the trust values of different companies and then will help to establish a connection between client and the company chosen by the client. Trust values will be calculated and stored in broker system. Companies need to inform broker system about the updated information of trust values (after updating by the last user) and broker system will update the trust related values of that company to its own website so that the next client can easily see the present condition of the company. E-business company will keep information about its lower level servers' and databases' which will be provided by the broker system so that company can use this information for rating its own servers' and databases' for providing best service to clients and users. Companies can keep their transaction information and invoices in different databases by encrypting them with AES algorithm and keep the keys (automatically generated by companies' main system computer) in the main system computer's database. The whole architecture is described below with specific points in the figure:

1. This model uses RSA algorithm for secured communication between client, main system (Broker) and different e-commerce related companies. For sending requests between broker and then commercial companies (which are considered using cloud structure), first of all, they need to communicate with the key distribution system for sending their public keys. With the help of private key, client/user will encrypt his information and request which he wants to send to broker system and then to the company's IT system whom he wants to deal with after getting information about previous trust values and behavioral probability of the company.

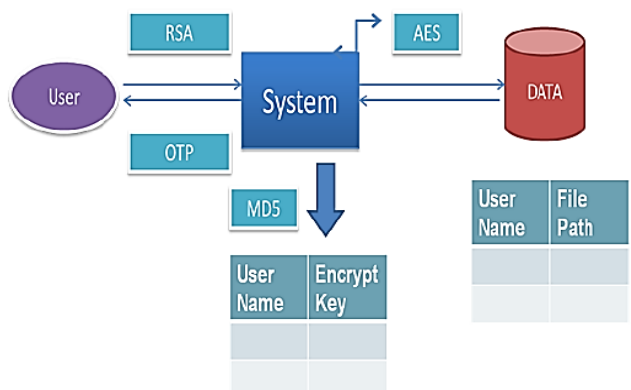

Figure 1. Security Architecture

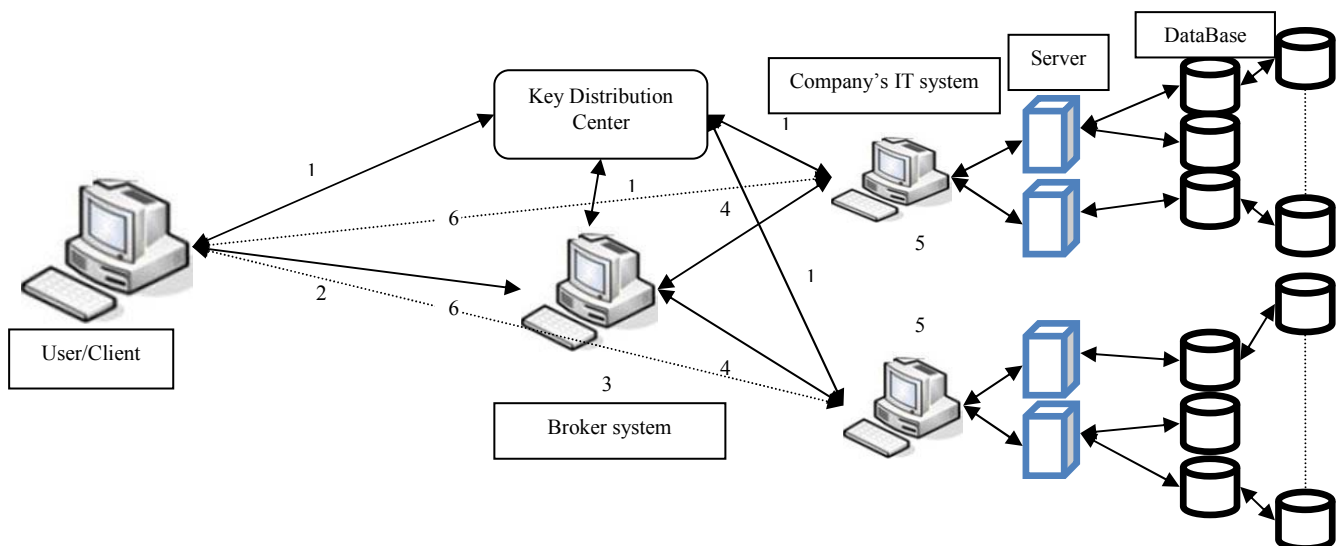

Figure 2: Proposed E-Commerce Architecture

2. Client needs to see the trust values and behavioral probability of e-business companies for searching out best company. Broker system will serve client all information about different companies. Client will choose one from them and then easily continue his/her transactions with that company via broker system.

3. Broker system will keep all the information related to Trust T and Behavioral Probability P of different e-business companies in their database. Broker system will also calculate Trust values and Behavioral probabilities for different companies by getting information of different servers of different companies. Broker system will be responsible for all types of information of different companies related to trust values and behavioral probabilities.

4. Broker system will communicate with company after selection of that company by the client and ask the company to communicate with client system. RSA encryption system is proposed to use between them and company's IT system will send the evaluation of servers' given by the client to broker system for calculating trust values of the company.

5. Every company will maintain his/her own servers and distributed database systems by a main PC. This system will keep the 128 bit keys which are used for encrypting different invoices and files for business between different clients and companies. It can also keep files on different topics in his own database or company's distributed database systems. This pc will maintain all the servers of the company and can rate its own servers and databases by evaluating its clients ratings.

6. After selecting one company from the broker webpage, the broker system will establish a peer to peer connection between the client and the company. It will help the clients to work freely, easily and securely complete their transaction with high rated speed. Every client will think that he is the only user of that company.

## V. EXPERIMENTAL RESULTS

In Lab, we had worked with the whole proposed e-commerce architecture for ensuring its validity and feasibility of working environments. We had got different results for our work. In case study 1 result of Trust model and in case study 2 result of security model are discussed. We had used following configuration during our experimental execution.

- Platform: Visual Studio 2010 (asp.net)
- Processor: Core 2 Duo (2.93 GHz),
- RAM: 2 GB

*A. Case Study 1:*

According to Certain Trust Model's operators defined in equation 3, 4, 5 and Table 1, we know that, the input for this model is r, s, f and w. Let, the input values are r=5, s=2, f=0.5 and w=1. Here, no of evidences are N=7. Then, the output values are:- average rating t = 0.714 and c=0.724. and then, E = 0.65. Now, for mapping it to our proposed model, we need to modify t. Here,

$$t' = t * \text{scale of rating} \qquad (6)$$

Usually, the scale of rating is 5. Now, the new average rating is $t = 0.714*5 = 3.57$. Then, the value of parameter Trust, $T = ((3.57*0.724)/5)*100 = 51.69\%$. As the range of trust varies from person to person, one can consider it under high trust region. From the result of E and T, we can see that, it is easier to understand the condition of system or server much better that the past. As it is represented in percentage form, the user can easily understand the evaluated value of trust. Now, the value of second parameter, behavioral probability, p = 3.39% and because of T>f, the system is now in the condition of showing 3.39% higher behavior than the initial expectation. As it is in higher condition, so, the hosting partner can easily take the decision to host in this server/system. This parameter will also be useful for the developers so that they can easily understand the present condition of the system. Figure 3 shows the output.

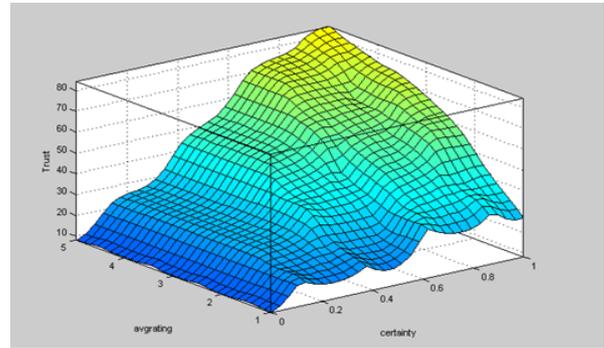

Figure 3: Mapping Surface of Fuzzified Certain Trust Model

*B. Case Study 2:*

Working with the model in Lab at different times and with different users and their individual commercial transactions, which are different from each other in size, contents, extension, prize amounts, etc. take different times for executing the overall model. Depending on the file size, program execution time varies from person to person. Among the 100 users result, 10 of them are shown in table II.

TABLE II. PTOCESSING TIME FOR WHOLE PROPOSED MODEL

| Person No | Transaction Time (Full Process) | Person No | Time Required for file Upload (Full Process) |
|---|---|---|---|
| 1 | 3 sec | 6 | 3 sec |
| 2 | 5 sec | 7 | 3 sec |
| 3 | 6 sec | 8 | 3.5 sec |
| 4 | 6.5 sec | 9 | 3 sec |
| 5 | 3 | 10 | 4 sec |

Let us consider two online business companies, A and B. After completing a number of few successful transactions with customers, Trust value for company A is 54% and behavioral probability is 9% upper than the base. On other side, company B's Trust value is 46%, whose behavioral probability is 6% lower than its base condition. Now, at this present state, when a new customer ask the broker for starting a transaction, the broker system will response to that customer with all information of both companies like figure 4. Now, customer will select one of the two companies and start his business.

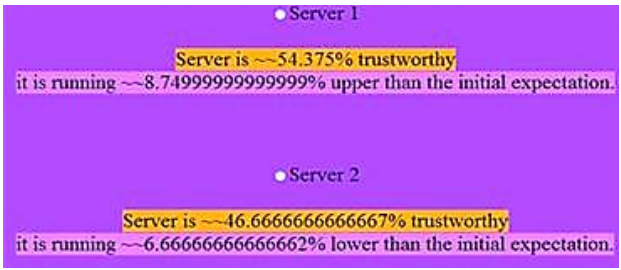

Figure 4: Information provided by the broker system

The advantages of our proposed model in the field of e-commerce are given below:

1. Users/clients can easily know the previous reputation of a business company from the broker webpage, so easily become more confident about his online transaction.
2. Companies can easily understand the present condition of their systems from the behavioral probability, so they can improve themselves easily.
3. Here, all the e-business companies need to run their business under a broker system. For these reason, they will try their best to remain stable in long run life of e-commerce competition.
4. Vandals and hackers can't easily engage in different types of hacking or create problems in e-commerce transactions because of high security and Trust value calculation system.

## VI. Conclusion

In this paper, we have proposed a new secured and trustworthy architecture for E-commerce which is fully cloud computing based. Present world is running towards cloud computing system. So, it is easy to imagine that future e-commerce will be depended on cloud computing. Different types of e-commerce systems are present now but most of them are individual system based. We have proposed a newer e-commerce architecture which is a combination of fuzzy based certain trust model and a highly secured encryption e-transaction based architecture. This architecture will help to solve present e-commerce problems related to cloud computing execution platform.

We have some future idea for working with this model. Firstly, we want to apply evolutionary algorithm with this model to optimize and design the rules. We want to apply price comparison as a parameter for a product in our model for ensuring accurate trust measuring model for a normal e-commerce website. Secondly, more development of behavioral probability parameter so that it can directly prohibit different types of security breaking questions like Sybil attack, false rating, etc is our fourth wish. Thirdly, In our proposed model we have used RSA encryption system which is deterministic. For this reason, it becomes fragile in long run process. But the other algorithms make the model highly secured. In future we want to work with ensuring secure communication system between users and system, user to user. We also want to work with encryption algorithms to find out more light and secure encryption system for secured file information preserving system.

ACKNOWLEDGMENT

Authors are cordially giving thanks to the researchers of different trust models, security models of cloud computing and all others who have tried hard to make their work easy to accomplish.